\documentclass[showpacs,superscriptaddress,amsmath,amssymb,reprint,prl]{revtex4-1}

\usepackage{graphicx}

\begin{document}

\title{From ultracold electrons to coherent soft x-rays}

\author{J.G.H. Franssen}
\affiliation{Department of Applied Physics, Eindhoven University of Technology, P.O. Box 513, 5600 MB Eindhoven, The Netherlands}
\affiliation{Institute for Complex Molecular Systems, Eindhoven University of Technology, P.O. Box 513, 5600 MB Eindhoven, The Netherlands}
\author{B.H. Schaap}
\affiliation{Department of Applied Physics, Eindhoven University of Technology, P.O. Box 513, 5600 MB Eindhoven, The Netherlands}
\author{T.C.H. de Raadt}
\affiliation{Department of Applied Physics, Eindhoven University of Technology, P.O. Box 513, 5600 MB Eindhoven, The Netherlands}
\author{D.F.J. Nijhof}
\affiliation{Department of Applied Physics, Eindhoven University of Technology, P.O. Box 513, 5600 MB Eindhoven, The Netherlands}
\author{P.H.A. Mutsears}
\affiliation{Department of Applied Physics, Eindhoven University of Technology, P.O. Box 513, 5600 MB Eindhoven, The Netherlands}

\author{O.J. Luiten}
\email{o.j.luiten@tue.nl}
\affiliation{Department of Applied Physics, Eindhoven University of Technology, P.O. Box 513, 5600 MB Eindhoven, The Netherlands}
\affiliation{Institute for Complex Molecular Systems, Eindhoven University of Technology, P.O. Box 513, 5600 MB Eindhoven, The Netherlands}

\date{\today}

\begin{abstract}%
Electromagnetic radiation in the soft x-ray spectral range ($1-100~\rm{nm}$ wavelengths or $0.01-1~\rm{keV}$ photon energies) is rapidly gaining importance in both fundamental research and industrial applications. At present the degree of coherence and the average photon flux required by advanced applications is only available at large-scale synchrotron facilities and Free Electron Lasers (FELs), severely limiting the range of applications. We propose a fully coherent soft x-ray source, based on Inverse Compton Scattering (ICS) of electron bunches created by photoionization of a laser-cooled and trapped atomic gas. By combining spatial modulation of the photoionization process with radiofrequency bunch compression techniques, micro-bunching at soft x-ray wavelengths and thus coherent amplification can be realised, resulting in a soft x-ray table-top Compton light source. 

\end{abstract}


\maketitle

X-ray methods are amongst the most powerful non-destructive tools for analysing matter and therefore of enormous importance to society. Fields like biomedicine, material science and crystallography rely heavily on the imaging and analytical capabilities of soft and hard x-rays. The conventional x-ray source in most research labs is the x-ray tube, which has three important limitations: relative low intensity, poor coherence and selective availability of x-ray energies. In contrast, synchrotrons and, even more, x-ray free electron lasers (XFELs) offer high-brilliance, coherent and energy-tunable x-rays. Yet, these machines are only available at specialised, large scale facilities, providing scarce beam time at high cost outside the scientists lab. In order to bridge the gap between x-ray tubes and large scale facilities, compact coherent x-ray sources with high brilliance are necessary. Several methods like discharge/laser produced plasmas or high harmonic generation from noble gases are promising table top solutions but lack either tunability, coherence or flux at high photon energies.

In this letter we will show that by using the low emittance, laser cooled ultra cold electron source (UCES) as injector for a ICS based light source will lead to a fully spatial coherent soft x-ray beam. In addition, the unique control of the photoionization process of laser cooled and trapped atoms will allow the production of a micro-bunched electron beam, leading to coherent amplification of the ICS process by super-radiant emission and thus a longitudinally coherent soft x-ray beam of unprecedented brilliance, which can only be achieved presently at large-scale XFEL facilities.

The physical basis is the ICS process~\cite{PhysRev.21.483}, in which part of a laser beam is bounced off a relativistic electron beam, turning it into a bright x-ray beam through the relativistic Doppler effect. If light with wavelength $\lambda_{0}$, coming in at an angle $\theta_{\rm{in}}$ with respect to the electron beam, is scattered into an angle $\theta_{\rm{out}}$, then the wavelength of the scattered light is given by:
\begin{equation}
\lambda_{\rm{x}}=\lambda_{0}\frac{~1-\beta \cos(\theta_{\rm{out}})}{1+\beta \cos(\theta_{\rm{in}})}\label{wavelengthx}
\end{equation}
where $\beta=\frac{v}{c}$ is the velocity of the electrons normalized to the speed of light. For example, for a head-on collision, $\theta_{\rm{in}}=0$, a laser wavelength $\lambda_{0}=500~\rm{nm}$ and a moderately relativistic electron beam with kinetic energy $U=2~\rm{MeV}$, i.e. $\beta=0.98$ and $\gamma=5$, soft x-rays will be generated at wavelengths as short as $\lambda_{x}=5~\rm{nm}$. The x-rays will be emitted in a cone with a half angle of $\sim \gamma^{-1}$ centred around the direction of the electron beam. The shortest wavelengths are being generated in the forward direction ($\theta_{x}=0$) while progressively longer wavelengths are emitted for increasing $\theta_{x}$. The intrinsic narrowband nature of an ICS based source, combined with its high degree of directionality and the straightforward way in which the x-ray wavelength can be tuned continuously by simply changing the electron beam energy, make it a very attractive method for generating soft x-rays. 

Unfortunately, however, the efficiency of the ICS process is very low. Assuming an electron bunch of $N_{e}$ electrons focussed to an rms spot size $\sigma_{\rm{e}}$ colliding with a laser pulse of $N_{0}$ photons focussed to an a rms spot size $\sigma_{\rm{L}}$ the number of x-ray photons $N_{x}$ produced is given by 
\begin{equation}
N_{x}=\frac{N_{0}}{2\pi \sigma_{\rm{L}}^2}\frac{\sigma_{T} N_{e}}{1+\left(\frac{\sigma_{\rm{e}}}{\sigma_{\rm{L}}}\right)^{2}}\label{Nxray}
\end{equation}
with $\sigma_{T} = 6.65~10^{-29}~\rm{m}^{2}$ the Thomson scattering cross section. Equation~(\ref{Nxray}) shows that the number of generated x-rays is only dependent on the peak photon density $\left(\frac{N_{o}}{2\pi \sigma_{\rm{L}}^2}\right)$ and the number of photons $N_{e}$ in the electron pulse when the electron beam focus is much tighter than the laser focus, i.e. $\sigma_{\rm{e}}<<\sigma_{\rm{L}}$. 

As an example, if $\lambda_{0} = 500~\rm{nm}$, $100~\rm{mJ}$ laser pulses focussed to an rms beam size of $\sigma_{\rm{L}}=10~\mu \rm{m}$ are collided with $100~\rm{pC}$ electron bunches at a repetition rate of $1~\rm{kHz}$, then an x-ray flux $\Phi_{x} \approx 2~10^{10}~\rm{s}^{-1}$ will be generated. This is an optimistic estimate, assuming state-of-the-art pulsed electron and laser beam technology, but it is still $2-3$ orders of magnitude below the desired flux for advanced imaging applications. Moreover, the bandwidth will be large, as photons scattered at all angles are used in the estimate, and the spatial coherence of the generated soft x-ray beam will be very small, $< 10^{-2}$ partial coherence, due to the inevitably large angular spread of the electron beam, associated with the finite emittance of a $100~\rm{pC}$ bunch.
 
\begin{figure}
\includegraphics[width=0.5\textwidth]{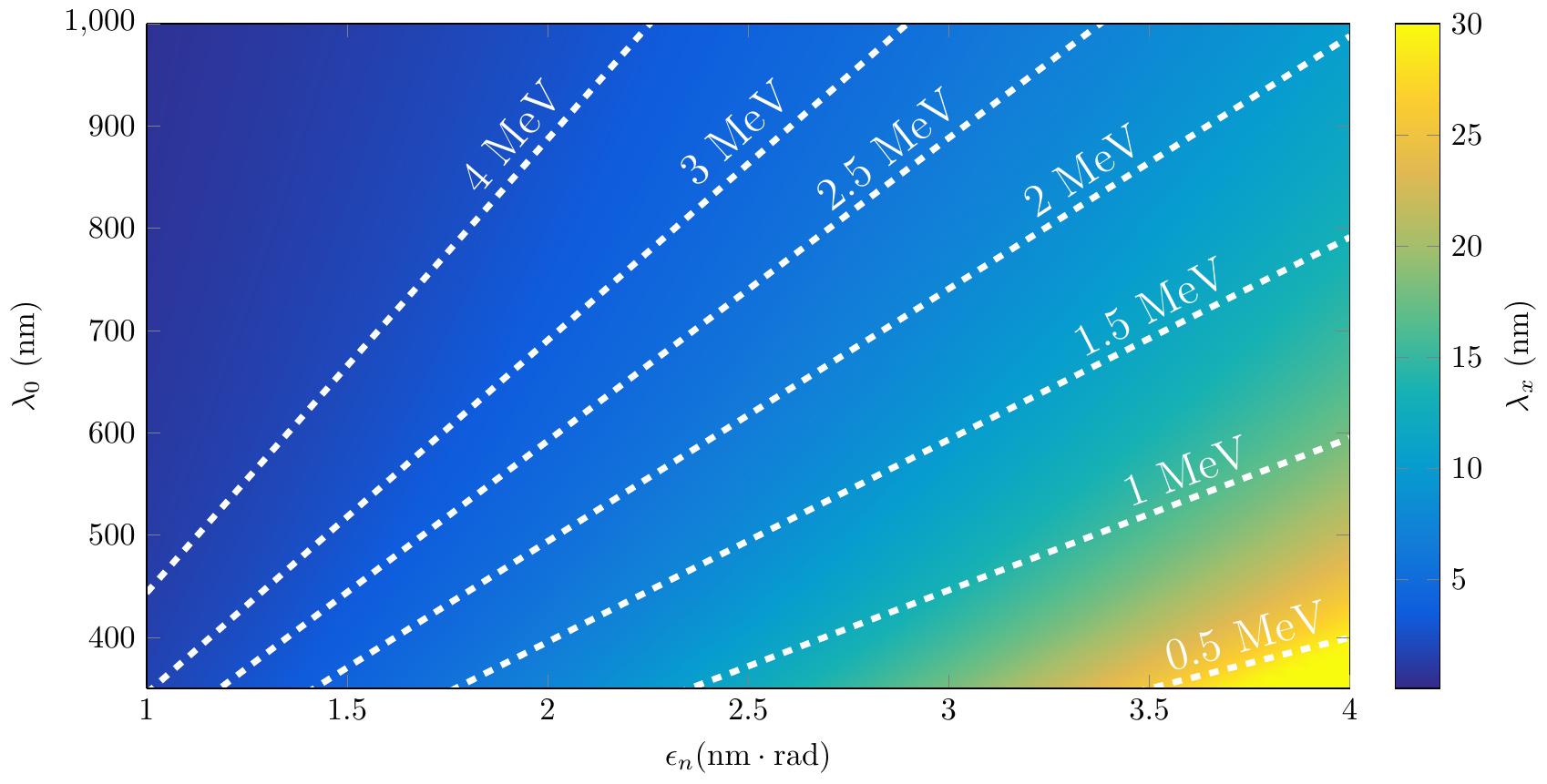}
\caption{\label{coherentICS}soft x-ray wavelengths $\lambda_{x}$ (color scale) that can be generated by spatially coherent ICS for given $\lambda_{0}$ and $\epsilon_{n}$. The required electron beam energies are indicated by the white dashed lines.}
\end{figure}

In order to generate a soft x-ray beam by ICS with full spatial coherence an electron beam with very high transverse quality is required. Transverse beam quality is usually expressed in terms of the geometrical emittance $\epsilon_{\rm{g}}$, or focusability of the beam, expressed in units $[\rm{m}~\rm{rad}]$, which is equal to the product of transverse beam size and uncorrelated angular spread. An electron beam can only generate a diffraction-limited, i.e. fully spatially coherent, x-ray beam if its emittance is much smaller than that of a diffraction limited x-ray beam, e.g. $\epsilon_{\rm{g}} < \frac{\lambda_{x}}{4\pi}$. Since geometrical emittance depends on beam energy, it is convenient to use the normalized emittance $\epsilon_{n}=\gamma \beta \epsilon_{g}$, which is a Lorentz invariant measure for beam quality. In terms of the normalized emittance the coherence condition becomes
\begin{equation}
\epsilon_{n}< \frac{\gamma \beta \lambda_{x}}{4\pi}.\label{coherentxray}
\end{equation}
By combining Eq.~(\ref{wavelengthx}) with $\theta_{0}=\theta_{x}=0$ and Eq.~(\ref{coherentxray}) with an equality sign, we can calculate the minimum conditions necessary for spatially coherent ICS, resulting in the colour plot shown in Fig.~\ref{coherentICS}.

Figure~\ref{coherentICS} shows the soft x-ray wavelengths $\lambda_{x}$ that can be generated by spatially coherent ICS for a given laser wavelength $\lambda_{0}$ and normalized emittance $\epsilon_{n}$. The required electron beam energy is indicated by the white dashed lines. For example, for $\lambda_{0}=500~\rm{nm}$, $\epsilon_{n}=3.4~\rm{nm}~\rm{rad}$ and $1~\rm{MeV}$ beam energy, spatially coherent soft x-ray radiation is generated with $\lambda_{x}=15~\rm{nm}$. It is immediately clear from Fig.~\ref{coherentICS} that in order to generate coherent soft x-ray radiation by ICS, high quality electron beams are required with normalized emittances preferably below $4~\rm{nm}~\rm{rad}$. Such beam qualities are usually associated with electron microscopy sources, which do not allow the generation of bunches with a lot of charge. 

In 2005 the ultracold electron source (UCES) was proposed~\cite{Claessens2005}, this source is a perfect match since it has the unique property of creating high charge electron pulses which can be shaped in 3D and which additionally have a large degree of transverse coherence. The UCES is based on near-threshold photo-ionization of a laser-cooled atomic gas in a magneto-optical trap (MOT). The UCES is characterised by electron temperatures as low as $10~\rm{K}$, $2-3$ orders of magnitude lower than conventional photoemission sources, as was demonstrated first by nanosecond photoionization~\cite{Claessens2007,Taban2010} and later by femtosecond photoionization as well~\cite{Engelen2013,McCulloch2013}. It is clear that the UCES allows much smaller normalized emittances than are possible with conventional photoemission sources. For example, for an rms transverse source size $\sigma_{s}=25~\mu\rm{m}$ and electron temperature $T=10~\rm{K}$, the normalized emittance $\epsilon_{n}=1~\rm{nm}~\rm{rad}$, a value that is routinely achieved with the UCES~\cite{Engelen2013,McCulloch2013,VanMourik2014a,Franssen2017}.

The size of the trapped gas cloud and thus the longitudinal size of the ionization volume is typically $1~\rm{mm}$ and the densities can be as high as a few $10^{18}~\rm{m}^{-3}$, implying that $N_{e}=10^{6}-10^{7}$ electrons can be created with $\epsilon_{n}=1~\rm{nm} ~\rm{rad}$. This combination of bunch charge and beam quality should enable, e.g., single-shot protein crystallography~\cite{Taban2010,Luiten2015,Speirs2015a}, which is one of the main driving forces behind the development of the UCES.

It thus follows (see Fig.~\ref{coherentICS}) that by using the UCES as an electron injector for an ICS source, fully spatially coherent radiation can be generated over the entire soft x-ray spectral range. This is a unique property of the UCES and by itself an intriguing property. However, the amount of x-ray photons generated with such bunches will be very modest (see Eq.~(\ref{Nxray})). Fortunately, the special characteristics of the UCES allow another trick to be played, which will both boost the photon yield enormously and take care of the longitudinal coherence as well.

The resonant two-step photoionization process, employing the combination of an excitation laser, tuned to an intermediate atomic level, and an ionization laser, exciting atoms from the intermediate state to the continuum, allows very precise control of the initial density distribution of the ionised gas: since atoms are only ionised in the region where the two laser beams overlap, the initial electron bunch distribution can be accurately tailored in 3D by modulating the beam profiles of the two lasers. This was demonstrated by the Scholten group at the University of Melbourne, who used a spatial light modulator (SLM) to shape the excitation laser beam and thus create electron bunches with intricate, almost arbitrary charge distributions, with the smallest sized structures only limited by the diffraction of the laser light~\cite{McCulloch2011}. The low temperature of the source turns out to be essential to maintain these intricate structures, which immediately get blurred due to random thermal motion of the electrons at higher source temperatures.

\begin{figure}
\includegraphics[width=0.5\textwidth]{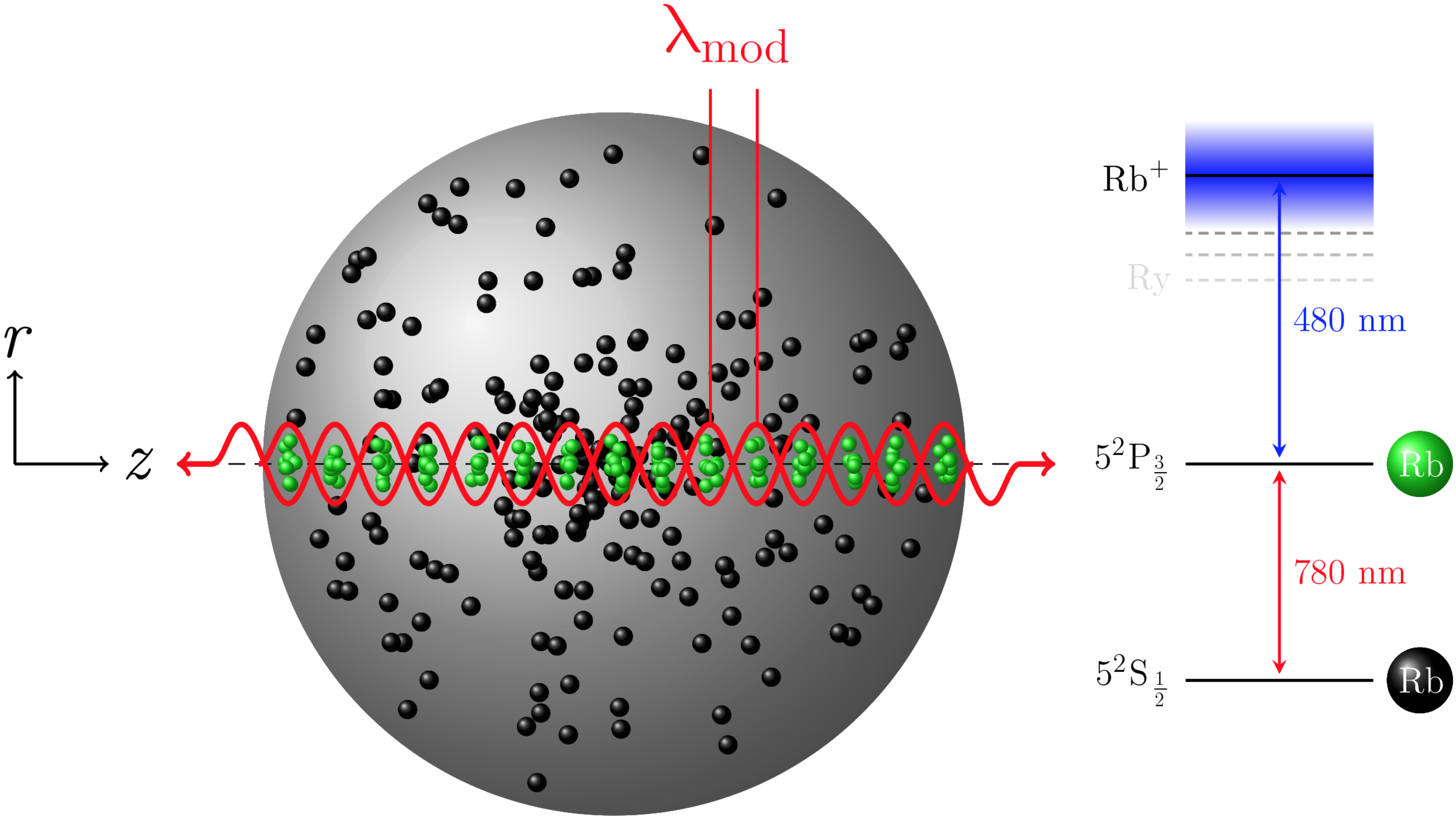}
\caption{\label{standingwaveRB} By using a $780~\rm{nm}$ standing wave (red) to excite the $5^{2}P_{3/2}$ state, the excited Rb gas in a MOT will be spatially modulated with a period of $\lambda_{\rm{mod}}$. Black spheres: laser-cooled ground state atoms; green spheres: atoms excited to the $5^{2}P_{3/2}$ state.}
\end{figure}

Here we propose a much simpler but very effective way to shape the initial charge distribution in a way extremely beneficial for boosting the ICS yield. As is illustrated in Fig.~\ref{standingwaveRB}, the excitation to the intermediate state can be done with two counter-propagating $780~\rm{nm}$ laser beams, creating a standing wave pattern. The periodically modulated excited gas is subsequently ionised by the $480~\rm{nm}$ femtosecond ionization laser, perpendicular to the excitation laser beam, thus almost instantly creating an electron bunch spatially modulated with a period equal to half the excitation laser wavelength, i.e. $\lambda_{\rm{mod}}=390~\rm{nm}$ (see Fig.~\ref{standingwaveRB}). This results in a longitudinal electron density modulation of the form $\sim \cos^{2}\left(\frac{\pi z}{\lambda_{\rm{mod}}}\right)$.

The structure is maintained in longitudinal phase space as long as the random thermal energy of the electrons $kT$ is much smaller than the kinetic energy difference gained between two succeeding micro-bunches $eF\lambda_{\rm{mod}}$, where $F$ the electric field strength of the  acceleration field. As an example, for $F=1~\rm{MV}/\rm{m}$ this implies that the electron temperature $T<<10^{4}~\rm{K}$ which is easily achievable using the UCES.

To generate soft x-ray radiation by ICS, the electron bunch has to be accelerated to the required energy, somewhere between $0.5$ and $4~\rm{MeV}$ (see Fig.~\ref{coherentICS}). This will require radio-frequency (RF) accelerator structures. Because of the high accelerating fields in the RF accelerators, typically $10~\rm{MV}/\rm{m}$, only a few cells, and thus less than half a meter of accelerator structure is sufficient to cover the entire soft x-ray spectral regime. Only acceleration, however, is not sufficient. In order to boost the ICS yield substantially coherent amplification is required. This can be accomplished by compressing the bunch in such a way that at the point where the accelerated bunch collides with the laser pulse, the period of the spatial modulation is decreased to the wavelength of the x-ray radiation generated. 

For example, by accelerating a bunch with a normalized emittance $\epsilon_{n}=3.4~\rm{nm}~\rm{rad}$ to an energy of $1~\rm{MeV}$ and colliding it with a $\lambda_{0}=500~\rm{nm}$ laser pulse, spatially coherent soft x-ray radiation is generated at a wavelength of $\lambda_{x}=15~\rm{nm}$ (see Fig.~\ref{coherentICS}). During initiation, the bunch is spatially modulated with a period equal to half the excitation laser wavelength, i.e. $\lambda_{\rm{mod}}=390~\rm{nm}$, so during acceleration the bunch has to be compressed by a factor $\frac{\lambda_{\rm{mod}}}{\lambda_{x}}\approx 20-30$ which is easily achievable using well established RF compression techniques~\cite{Oudheusden20101}. As a result, the fields of the radiation emitted by the individual micro-bunches will add up in phase, thus coherently amplifying the x-ray photon yield proportional to the bunch charge squared. Strictly speaking, coherent stimulated emission is added to the incoherent spontaneous emission $N_{X0}$, described by Eq.~(\ref{Nxray}), which results in~\cite{Carlsten} $N_{x} = (1+FN_{e})N_{X0}$, where $0 \leq F \leq 1$ is the form factor associated with the electron bunch distribution: in absence of any density modulation $F=0$, while $F=1$ for a bunch with a perfect periodic longitudinal density distribution. In our case we start with $F=0.25$, which is based on the modulation presented in Fig.~\ref{standingwaveRB}.

For bunch charges of $0.1~\rm{pC}$, i.e. $N_{e}=6.2~10^{5}$ electrons, colliding with $100~\rm{mJ}$, $\lambda_{0}=500~\rm{nm}$ laser pulses in a $\sigma_{\rm{L}}=10~\mu\rm{m}$ waist at a repetition rate of $1~\rm{kHz}$, the incoherent ICS photon flux (Eq.~(\ref{Nxray})) is $\Phi_{x}\approx 2~10^{7}~s^{-1}$. Assuming that we can maintain the $F=0.25$ density modulation, the coherent photon flux becomes $\Phi_{x}\approx 2.5~10^{12}~s^{-1}$, more than sufficient for advanced imaging techniques. To obtain the same photon flux by incoherent ICS would require focusing a sub-ps, few MeV, $60~\rm{nC}$ electron bunch to a spot smaller than $10~\mu\rm{m}$, which is not possible. In the calculation above a $33~\mu\rm{W}$ x-ray beam is generated by colliding the intense laser pulse with an electron bunch that is carrying only $100~\mu\rm{W}$ of beam energy. Compton scattering in the Thompson limit assumes that the fraction of momentum transferred from the electron to the x-ray photon is much smaller than unity. Therefore, $\Phi_{x}=1~10^{11}~s^{-1}$ is a more realistic value since now each electron only loses a small fraction $(4\%)$ of its energy. This limitation, however, can be circumvented in several ways, e.g. by increasing the electron beam energy while simultaneously increasing the laser wavelength $\lambda_{0}$, or alternatively by simply increasing the bunch charge.

\begin{figure*}
\includegraphics[width=0.9\textwidth]{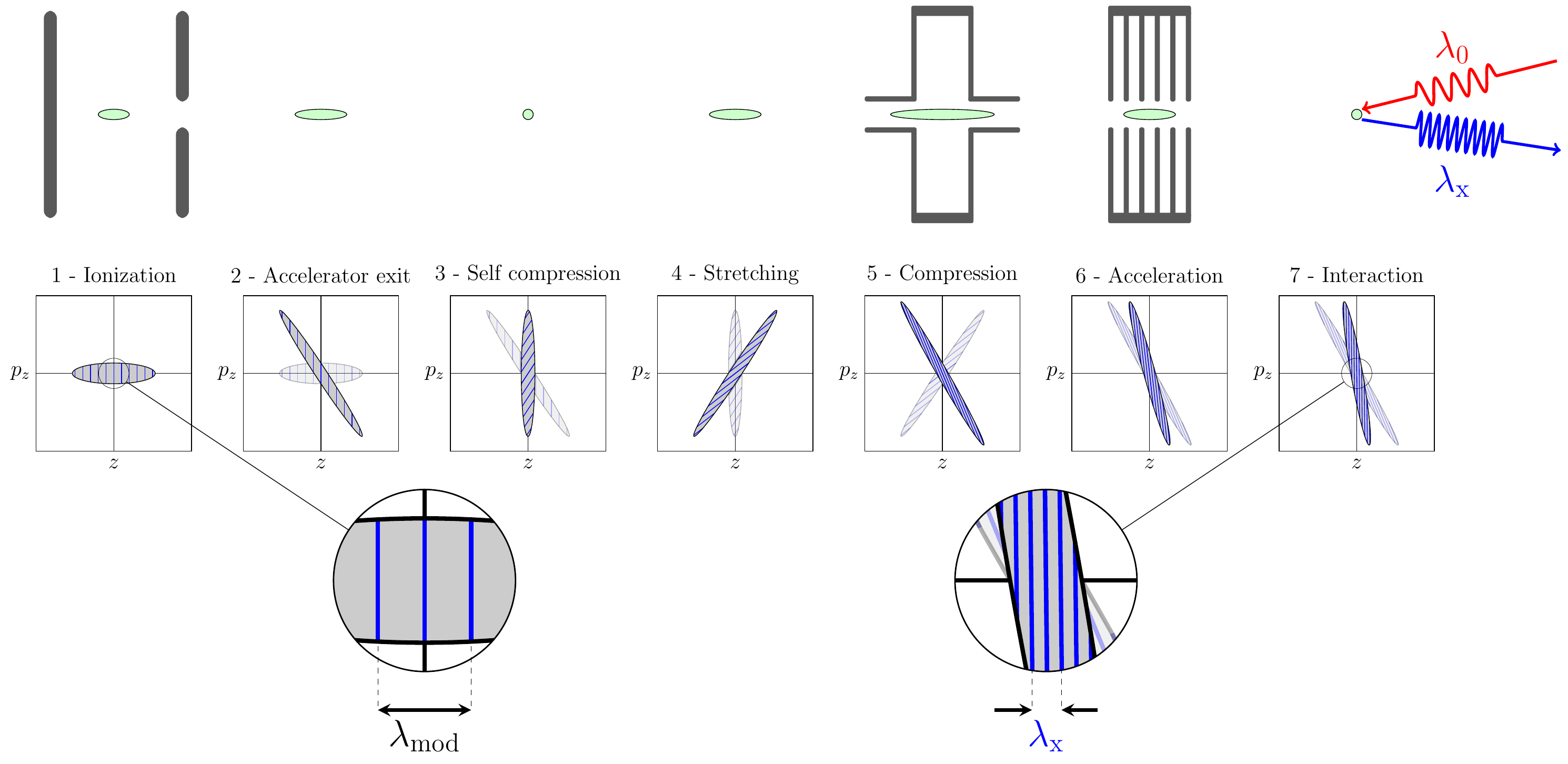}
\caption{Principle of RF bunch compression (not to scale). Top: electron bunch propagating through beam elements. From left to right: (1) ionization inside the grating-MOT-based UCES; (2) after exiting the DC accelerator the bunch has acquired a negative energy chirp, leading to velocity bunching; (3) self-compression point; (4) positive energy chirp and thus stretching of the bunch; (5) RF compression cavity inverting the chirp; (6) X-band accelerator section boosting the average bunch energy; (7) interaction point. Bottom: schematic illustration of the evolution of the longitudinal phase space distribution.}
\label{bunchcompression}
\end{figure*}

The coherent amplification of pulsed-electron-beam based radiation sources by this so-called super-radiance mechanism is well known and has been applied several times before. The challenge is always how to realise the required longitudinal density modulation, in the case of soft x-ray radiation at the nanometer scale. Already in 1996 Carlsten et al. proposed to apply the density modulation in the transverse direction first, which can be done quite straightforwardly with a mask, and subsequently use a magnetic chicane to transfer it to the longitudinal direction~\cite{Carlsten}. The Graves group at MIT/ASU has recently devised a particularly smart variation of this method to actually realise nano-modulated electron beams and thus use super-radiance to coherently amplify the soft x-ray photon yield in an ICS setup~\cite{Graves2018}. The UCES based method we propose here, has two major advantages: first, the two-step photoionization method allows extremely accurate shaping of the initial longitudinal bunch density distribution (see Fig.~\ref{standingwaveRB}); second, the UCES based method provides full spatial coherence.

The combination of superradiant amplification of the emission by micro-bunching of the electron bunch and fully spatially coherent emission, constitutes the realisation of a fully coherent soft x-ray source. The UCES-based soft x-ray source would have a footprint of only a few square meters, in stark contrast with present-day FEL facilities. Clearly, this would be an enormously important development allowing wide-spread dissemination of soft x-ray FELs in academic and industrial labs and potentially even in semiconductor fabs.

Although in principle the UCES provides the ingredients necessary to realise full spatial coherence and superradiant emission, there are still major obstacles facing actual realisation of a soft x-ray Compton FEL. These obstacles can be summarised in a single, major challenge: the control of space charge forces. To achieve a large photon flux, as many electrons as possible should radiate in perfect unison, while confined in a very small volume, both focused transversely to a few $\mu\rm{m}$, and compressed longitudinally to a few $10~\mu\rm{m}$ (temporal compression to $\sim 100~\rm{fs}$). The space charge forces associated with these high charge densities could cause deformation of the phase space distribution of the bunch, which could lead to irreversible emittance growth and thus loss of spatial coherence. Moreover, space charge forces could hamper bunch compression, leading to a deteriorated density modulation at the interaction point and thus reduced super-radiance.

In Fig.~\ref{bunchcompression} the principle of RF bunch compression is explained in a schematic illustration of a possible realisation of the UCES-based ICS setup we propose. The electron bunches are created inside the grating-MOT-based UCES (1) with a longitudinal periodic density modulation, schematically indicated by vertical blue bars. The electrons are accelerated to a few $10~\rm{keV}$ in an electrostatic accelerator. Because the electrons created in the back are accelerated over a larger distance, they acquire a larger kinetic energy and thus a higher velocity than those in front. As result the particles in the back will overtake the ones in front, leading to a self-compression point (3) due to velocity bunching just outside the UCES. This self-compression point is too close to the UCES to be useful for ICS (schematic drawing is not to scale). After the self-compression point the bunch gets stretched as the distribution acquires a positive energy chirp (4). The electrons are subsequently injected into a $3~\rm{GHz}$ resonant RF cavity in $\rm{TM}_{010}$-mode (5), in which the chirp is inverted, acquiring a strong negative chirp again, leading to bunch compression by velocity bunching in the drift space behind the RF cavity. While contracting, the bunches enter a RF accelerator section (6) in which they are accelerated to the desired energy. After the accelerator section the bunches reach the second compression point. Just before maximum compression, exactly at the point where the density modulation (indicated by blue bars) is properly lined up again (7), the bunches collide with the ICS laser, generating a soft x-ray beam. Preliminary charged particle tracking simulations using the GPT code~\cite{GPT} show that the required density modulation can indeed be achieved in the interaction point, with conservation of emittance.

The periodic spatial modulation in the MOT could also be accomplished in the ground state gas by using the dipole force in the standing wave of two counter propagating laser beam at a wavelength far-detuned to the blue with respect to the transition to the intermediate state. In fact, this could be a superior method, since the it would entail compressing the atoms prior to excitation, thus leading to higher initial bunch densities.

We propose a soft x-ray source, based on Inverse Compton Scattering of electron bunches created by photoionization of a laser-cooled and trapped atomic gas. By combining spatial modulation of the photoionization process with radiofrequency bunch compression techniques, micro-bunching at soft x-ray wavelengths and thus coherent amplification can be realised. Additionally, the low transverse electron beam emittance of the ultracold electron source allows for the generation of a fully spatial coherent x-ray beam.

\begin{acknowledgments}

This research is supported by the Institute of Complex Molecular Systems (ICMS) at Eindhoven University of Technology. 

\end{acknowledgments}

\bibliography{library.bib}

\end{document}